\documentclass[11pt]{article}
\pdfoutput=1
\usepackage{jheppub}
\usepackage{amssymb,amsfonts,accents}
\usepackage{mathrsfs}
\let\savenumberline\numberline
\def\numberline#1{\savenumberline{#1.}}
\makeatletter
\renewcommand{\@seccntformat}[1]{\csname the#1\endcsname.\,\,}
\makeatother

\newcommand{\C}{{\bf C}}

\newcommand{\CA}{{\cal A}}

\newcommand{\CF}{{\cal F}}

\newcommand{\CO}{{\cal O}}

\newcommand{\CZ}{{\cal Z}}

\newcommand{\SC}{{\mathscr C}}
\newcommand{\SD}{{\mathscr D}}

\newcommand{\SJ}{{\mathscr J}}
\newcommand{\SM}{{\mathscr M}}

\newcommand{\SZ}{{\mathscr Z}}

\renewcommand{\hat}[1]{\widehat{#1}}
\newcommand{\be}{\begin{equation}}
\newcommand{\ee}{\end{equation}}
\newcommand{\bea}{\begin{eqnarray}}
\newcommand{\eea}{\end{eqnarray}}

\newcommand{\ie}{\textit{i.e.}}
\newcommand{\eg}{\textit{e.g.}}

\newcommand{\Tr}{\textrm{Tr}}
\newcommand{\sigp}{{\Sigma^+}}
\newcommand{\sigm}{{\Sigma^-}}
\newcommand{\sigw}{{\Sigma^\wedge}}
\newcommand{\cl}{\textrm{cl}}
\newcommand{\qu}{\textrm{qu}}
\newcommand{\rmm}{\textrm{M}}

\makeatletter
\def\@fpheader{\relax}
\makeatother
\usepackage{graphicx}
\usepackage{latexsym}

\title{\ \vspace{1.5in} \\ Keldysh Rotation in the Large-N Expansion\\ and String Theory Out of Equilibrium}
\author{Petr Ho\v{r}ava and Christopher J. Mogni}
\affiliation{Berkeley Center for Theoretical Physics and Department of Physics\\
University of California, Berkeley, CA, 94720-7300, USA\medskip\\
Theoretical Physics Group, Lawrence Berkeley National Laboratory\\
Berkeley, CA 94720-8162, USA}
\abstract{We extend our study of the large-$N$ expansion of general non-equilibrium many-body systems with matrix degrees of freedom $M$, and its dual description as a sum over surface topologies in a dual string theory, to the Keldysh-rotated version of the Schwinger-Keldysh formalism.  The Keldysh rotation trades the original fields $M_\pm$ -- defined as the values of $M$ on the forward and backward segments of the closed time contour -- for their linear combinations $M_{\textrm{cl}}$ and $M_{\textrm{qu}}$, known as the ``classical'' and ``quantum'' fields.  First we develop a novel ``signpost'' notation for non-equilibrium Feynman diagrams in the Keldysh-rotated form, which simplifies the analysis considerably.  Before the Keldysh rotation, each worldsheet surface $\Sigma$ in the dual string theory expansion was found to exhibit a triple decomposition into the parts $\Sigma^\pm$ corresponding to the forward and backward segments of the closed time contour, and $\Sigma^\wedge$ which corresponds to the instant in time where the two segments meet.  After the Keldysh rotation, we find that the worldsheet surface $\Sigma$ of the dual string theory undergoes a very different natural decomposition: $\Sigma$ consists of a ``classical'' part $\Sigma^{\textrm{cl}}$, and a ``quantum embellishment'' part $\Sigma^{\textrm{qu}}$.  We show that both parts of $\Sigma$ carry their own independent genus expansion.  The non-equilibrium sum over worldsheet topologies is naturally refined into a sum over the double decomposition of each $\Sigma$ into its classical and quantum part.  We apply this picture to the classical limits of the quantum non-equilibrium system (with or without interactions with a thermal bath), and find that in these limits, the dual string perturbation theory expansion reduces to its appropriately defined classical limit.}
\begin{document}
\maketitle
\section{Introduction}
\label{ssint}

In our previous paper \cite{neq}, to which this paper is a sequel, we studied the structure of the large-$N$ expansion of non-equilibrium systems with matrix degrees of freedom using the Schwinger-Keldysh formalism,%
\footnote{See \cite{neq} for an extensive list of references on the Schwinger-Keldysh formalism and some applications.}
and its dual description in terms of strings.  The goal of \cite{neq} was to use this duality to identify some of the first elements of the universal calculus for non-equilibrium string perturbation theory.

We limited our attention in \cite{neq} to the ``forward-backward'' (henceforth referred to as ``$\pm$'') representation of the Schwinger-Keldysh formalism: The system is evolved along an oriented closed time contour $\SC$, which consists of a forward component $C_+$ evolving from an early time $t_0$ to a late time $t_\wedge$, followed by the backward component $C_-$ going back from $t_\wedge$ to $t_0$.  This leads to a doubling of fields as functions of the single coordinate time $t$: For each field $\phi$, we denote by $\phi_+(t)$ the values of $\phi$ on $C_+$, and by $\phi_-(t)$ the values of $\phi$ on $C_-$.   

It is well known (see for example \cite{vilkovisky,rammer,kamenev,kamenevlesh,kamenevre}) that many important physical features of the Schwinger-Keldysh formalism for non-equilibrium systems are revealed in a different representation, involving a simple but very useful field redefinition.  Instead of the $\phi_\pm$ fields, this representation uses their sum and difference,
\bea
\phi_\cl&=&\frac{1}{2}(\phi_++\phi_-),
\label{eekelc}\\
\phi_\qu&=&\phi_+-\phi_-.
\label{eekelq}
\eea
The variables $\phi_\cl$ and $\phi_\qu$ are often referred to as ``classical'' and ``quantum'' \cite{kamenev,kamenevlesh,kamenevre}, even though they of course both represent fluctuating fields.  This field redefinition is usually referred to as the ``Keldysh rotation,'' since its idea goes back to \cite{keldysh}.  It leads to remarkable simplifications.  First, in the $\pm$ formalism, there are four nonzero propagators $G_{\pm\pm}$ satisfying one sum-rule identity
\be
\label{eesum}
G_{++}+G_{--}=G_{+-}+G_{-+}.  
\ee
The implications of this identity are often obscure in individual Feynman diagrams.  After the Keldysh rotation, only three propagators are nonzero:
\bea
\left\langle\phi_\qu(t')\,\phi_\cl(t)\right\rangle_0&\equiv&G_A(t',t),\\
\left\langle\phi_\cl(t')\,\phi_\qu(t)\right\rangle_0&\equiv&G_R(t',t),\\
\left\langle\phi_\cl(t')\,\phi_\cl(t)\right\rangle_0&\equiv&G_K(t',t),\\
\left\langle\phi_\qu(t')\,\phi_\qu(t)\right\rangle_0&\equiv&0.
\eea
Thus, in the Keldysh-rotated basis, the sum rule equivalent to (\ref{eesum}) is automatically satisfied, reducing the number of diagrams that need to be summed.   The second -- and physically more important -- simplification is that in the Keldysh basis, the information about the dynamics and the information about the state have been decoupled from each other:  The mixed propagators $G_A$ and $G_R$ are state-independent, and the entire information about the state is carried by $G_K$.  In contrast, in the $\pm$ formalism all four propagators $G_{\pm\pm}$ are sensitive to both the dynamics and the state.  These features of the Keldysh formalism make not only practical calculations more efficient, but also the physical picture more direct and easier to interpret.

To illustrate this well-known usefulness of the Keldysh rotation, consider the example of a relativistic scalar field of mass $m$ in thermal equilibrium at temperature $T$.  In the $\pm$ formalism, the momentum-space propagators (in the mostly-minus spacetime metric signature) are
\bea
G_{++}(p)&=&\frac{i}{p^2-m^2+i\epsilon}+2\pi\,n_{\textrm{B}}(|p^0|)\,\delta(p^2-m^2),\nonumber\\
G_{--}(p)&=&\frac{-i}{p^2-m^2-i\epsilon}+2\pi\,n_{\textrm{B}}(|p^0|)\,\delta(p^2-m^2),\nonumber\\
G_{+-}(p)&=&2\pi\left[\theta(-p^0)+n_{\textrm{B}}(|p^0|)\right]\delta(p^2-m^2),\vphantom{\frac{i}{m^2}}\nonumber\\
G_{-+}(p)&=&2\pi\left[\theta(p^0)+n_{\textrm{B}}(|p^0|)\right]\delta(p^2-m^2)\vphantom{\frac{i}{m^2}},\nonumber
\eea
where
\be
n_{\textrm{B}}(\omega)=\frac{1}{\exp(\omega/T)-1}
\ee
is the Bose-Einstein distribution function.  After the Keldysh rotation, we get just three nonzero propagators,
\bea
\left\langle\phi_\qu\,\phi_\cl\right\rangle_0&=&\frac{i}{p^2-m^2+i\,\textrm{sign}(p^0)\epsilon}\equiv G_A(p),\\
\left\langle\phi_\cl\,\phi_\qu\right\rangle_0&=&\frac{i}{p^2-m^2-i\,\textrm{sign}(p^0)\epsilon}\equiv G_R(p),\\
\left\langle\phi_\cl\,\phi_\cl\right\rangle_0&=&2\pi\left[\mbox{$\frac{1}{2}$}+n_{\textrm{B}}(|p^0|)\right]\delta(p^2-m^2)=\pi\coth\left(\frac{|p^0|}{2T}\right)\delta(p^2-m^2)\equiv G_K(p).
\eea
As promised, the quantum-to-quantum propagator vanishes identically, the mixed propagators become the advanced and retarded propagators $G_A$ and $G_R$ which only know about the dynamics but not about the state, and all the information about the initial density matrix is stored in the classical-to-classical propagator $G_K$.  

The Keldysh rotation also has a number of closely related cousins, which appear across a multitude of diverse areas of physics, always with similar simplifying results. In the Larkin-Ovchinnikov representation \cite{lo}, popular in non-equilibrium condensed matter \cite{rammer,kamenev}, another unitary transformation is performed on the fields; the same three propagators appear, but the propagator $2\times 2$ matrix is now upper triangular, with $G_A$ and $G_R$ on the diagonal, and $G_K$ in the upper-right corner.  In the closely related Langreth-Wilkins representation \cite{lw} (see \cite{spicka} for a review), popularized by the influential lecture \cite{langreth} and now wide-spread in use in non-equilibrium physics of mesoscopic systems \cite{svl}, a non-unitary field transformation is performed such that the propagator matrix stays upper triangular as in the Larkin-Ovchinnikov representation, but with the $G_K$ propagator replaced by $G^<$.  The Keldysh rotation (\ref{eekelc}) and (\ref{eekelq}) also plays a prominent role in the theory of the decoherence functional of Gell-Mann and Hartle \cite{ghh},%
\footnote{In this context, the rotation acts on two alternative histories $\phi_+(t)$ and $\phi_-(t)$ of the system that enter the decoherence functional; see also \cite{fv,fhibbs} for the earlier and closely related concept of an influence functional.}
which is instrumental in the description of the quantum-to-classical transition in the sum-over-histories approach to quantum systems, including those involving the dynamical spacetime geometries of quantum gravity and cosmology.  In this paper, we will concentrate on the original Keldysh rotation in its original context, but we expect that our results can be extended straightforwardly to such closely related cases as well.

In \cite{neq}, we used the $\pm$ version of the Schwinger-Keldysh formalism to derive some universal implications of the large-$N$ expansion for the dual string theory.  We found that, in comparison to strings at equilibrium, the string perturbation expansion is further refined, with each worldsheet $\Sigma$ subdivided into a triple decomposition,
\be
\Sigma=\sigp\cup\sigw\cup\sigm.
\label{eetriple}
\ee
Here the forward part $\Sigma^+$ is associated with the forward part $C_+$ of the time contour, and similarly for the backward part $\Sigma^-$ and $C_-$.  The ``end of time'' wedge region $\Sigma^\wedge$ is associated with the meeting point of $C_+$ and $C_-$ at $t_\wedge$, and it provides a bridge between $\sigp$ and $\sigm$.  Remarkably, each of the three parts of this triple decomposition of $\Sigma$ has its own associated genus expansion.

In view of the importance of the Keldysh-rotated version of the Schwinger-Keldysh formalism, in this paper we extend our analysis of the large-$N$ expansion and string theory to this Keldysh-rotated case.  Our results were briefly announced in \cite{ssk}, which also contains a brief summary of the results of \cite{neq} in the $\pm$ formalism.  Here we provide our detailed arguments and proofs justifying the statements announced in \cite{ssk}, and we also present additional results not advertised in \cite{ssk}.  Our analysis of the large-$N$ expansion again reveals an intriguing refinement of string worldsheet diagrams.  This time, however, the subdivision of worldsheets is not into the tree parts as observed in the $\pm$ formalism  -- instead, we will find a subdivision distinct from (\ref{eetriple}):
\be
\Sigma=\Sigma^\cl\cup\Sigma^\qu,
\ee
with the worldsheet $\Sigma$ composed of a ``classical'' part $\Sigma^\cl$, and its ``quantum embellishments'' $\Sigma^\qu$.  Each of the two parts of $\Sigma$ is again associated with its own genus expansion.  

The resulting picture of non-equilibrium string perturbation theory that emerges in the Keldysh-rotated formalism is by no means a straightforward consequence of the worldsheet picture established in \cite{neq} in the $\pm$ formalism based on the triple decomposition (\ref{eetriple}).  This is not entirely unexpected: Whereas in the language of the original matrix degrees of freedom, the Keldysh rotation is a rather simple change of variables, the worldsheet dual theories before and after the rotation should not be related in any simple way, for the following reason.  On the side of the matrix degrees of freedom, the Keldysh rotation mixes the values of $M$ on the forward and backward branches $C_+$ and $C_-$ of the time contour for the same value of $t$.  The simplicity of this mixing relies crucially on the existence of a canonical identification of the time evolution parameter $t$ along $C_+$ and $C_-$.  In contrast, things are not this simple on the worldsheets:  Even in the absence of knowing any details of the worldsheet dynamics, we anticipate some form of worldsheet diffeomorphism invariance, which makes any identification of the worldsheet time coordinate $\tau$ on $\sigp$ and $\sigm$ non-canonical at best, and impossible globally if $\sigp$ and $\sigm$ are of different topology (which they typically are).  The worldsheet representations before and after the Keldysh rotation will be related by a complicated resummation of many ribbon diagrams, and for these reasons, we do not anticipate any simple procedure for deriving one worldsheet picture from the other.  This is indeed the perspective supported by the main results of this paper.

\section{Large-\textit{N} expansion after the Keldysh rotation}

As in \cite{neq}, we start with a theory of $N\times N$ Hermitian matrix degrees of freedom $M^a{}_b(t,\ldots)$, which may be spacetime fields, or just quantum mechanical degrees of freedom; we only display the dependence on time, with the dependence on space and possible other quantum numbers playing only a spectator role in our arguments and therefore kept implicit.  In this way, our results will be universal, in particular independent of whether the theory is relativistic or not.  We further assume that the theory has an $SU(N)$ symmetry, and that the original action of the theory takes the single-trace form,
\be
S(M)=\frac{1}{g^2}\int dt\,\Tr\left(\dot M^2+M^3+M^4+\ldots\right).
\ee
We studied this theory on the Schwinger-Keldysh time contour $\SC$ in detail in Section~2 of \cite{neq}, and analyzed its large-$N$ expansion with the fixed 't~Hooft coupling $\lambda\equiv g^2N$, using the $\pm$ formalism.%
\footnote{As in \cite{neq}, it would be easy to generalize all our arguments to the case of more than one independent 't~Hooft coupling, controlling different interaction terms in $S(M)$.  We concentrate on one $\lambda$ for simplicity.  Also, as in \cite{neq}, we keep the dependence on spatial coordinates and spatial derivatives in the action implicit.}
In the $\pm$ formalism, $M(t)$ becomes doubled to $M_\pm(t)$, and the action is formally of the form
\be
S_{\textrm{SK}}(M_\pm)=S(M_+)-S(M_-),
\label{eeska}
\ee
which needs to be augmented by the appropriate boundary conditions:  the correct rules at the meeting point $t_\wedge$ between the two branches of $\SC$, and the information about the initial state at $t_0$ if different from the vacuum.  

Now we perform the Keldysh rotation of the fields: As in (\ref{eekelc}) and (\ref{eekelq}),  we define $M_\cl$ and $M_\qu$.  Each of these fields continues to carry the adjoint representation of our symmetry group.  In order to avoid notational clutter and too many subscripts and superscripts, we will use $M$ to denote the ``classical'' matrix field $M_\cl$,
\be
M(t)=\frac{1}{2}\left( M_+(t)+M_-(t)\right),
\ee
and $\SM$ to denote the ``quantum'' matrix field $M_\qu$:
\be
\SM(t)= M_+(t)-M_-(t).
\ee
After the Keldysh rotation, the action $S_{\textrm{SK}}$ becomes
\be
S_{\textrm{SK}}=\frac{1}{g^2}\int dt\,\Tr\left(K(M,\SM)+3\,M^2\SM+\frac{1}{4}\SM^3+4\,M^3\SM+\,M\SM^3+\ldots \right).
\label{eeactk}
\ee
The structure of the quadratic kinetic term $K(M,\SM)$ is such that it gives the three propagators that we discussed in Section~\ref{ssint}, as we will see again when we look at the Feynman rules below.  This form of the action would naturally generalize if we added higher polynomial interactions to $S(M)$, or allowed independent couplings to control different terms in $S(M)$.  Note, however, that the number of $\SM$'s in each monomial interaction term in (\ref{eeactk}) will always be odd.  

\subsection{Feynman rules for the ribbon diagrams after Keldysh rotation}

Feynman rules for the ribbon diagrams after the Keldysh rotation are as follows. The quadratic kinetic term $K(M,\SM)$ in (\ref{eeactk}) yields three propagators,%
\footnote{In our notation, we use the subscript ``$0$'' in $\langle\ldots\rangle_0$ to distinguish the bare propagators from the full 2-point functions $\langle\ldots\rangle$ which we will be studying below.}
\bea
\vcenter{\hbox{\includegraphics[width=.9in]{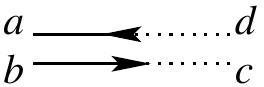}}}\ &=&\left\langle M^a_{\ b}\,\SM^c_{\ d}\right\rangle_0,
\label{eeprpms}\\
\vcenter{\hbox{\includegraphics[width=.9in]{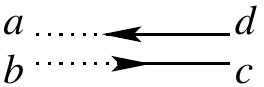}}}\ &=&\left\langle \SM^a_{\ b}\,M^c_{\ d}\right\rangle_0,
\label{eeprpsm}\\
\vcenter{\hbox{\includegraphics[width=.9in]{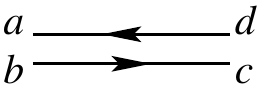}}}\ &=&\left\langle M^a_{\ b}\,M^c_{\ d}\right\rangle_0.
\label{eeprpmm}
\eea
We use the notation popular in the non-equilibrium field theory literature (see, \eg , \cite{kamenev}): The dotted line denotes the ``quantum'' end of a propagator, and the full line denotes the ``classical'' end.  Here we have just extended this convention to ribbons.

The vertices are
\bea
\label{eev3c}
\vcenter{\hbox{\includegraphics[width=.75in]{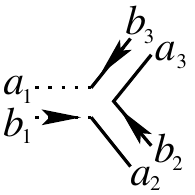}}}\ \ \ &=&\ \ \frac{N}{\lambda}\left(\ldots \right),\\
\label{eev3q}
\vcenter{\hbox{\includegraphics[width=.75in]{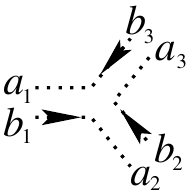}}}\ \ \ &=&\ \ \frac{N}{\lambda}\left(\ldots \right)
\eea
at three points, and
\bea
\label{eev4c}
\vcenter{\hbox{\includegraphics[width=.8in]{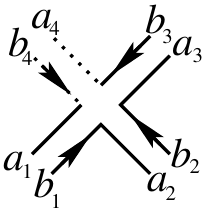}}}\ \ \ &=&\ \ \frac{N}{\lambda}\left(\ldots \right),\\
\label{eev4q}
\vcenter{\hbox{\includegraphics[width=.8in]{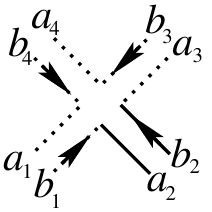}}}\ \ \ &=&\ \ \frac{N}{\lambda}\left(\ldots \right),\\
\vdots\qquad\quad\ \ &&\nonumber
\eea
at four points.  As in \cite{neq}, the vertical dots at the end of this list stand for higher $n$-point vertices, which we allow to be present for full generality, but do not depict explicitly.  Note that they all have to satisfy one restriction:  The number of quantum ends at each vertex always has to be odd, a feature that follows from the general structure of (\ref{eeactk}).  

The precise numerical values of the vertices can be easily extracted from (\ref{eeactk}) (or appropriate generalizations thereof).  The horizontal dots ``$(\ldots)$'' on the right-hand sides of (\ref{eev3c}-\ref{eev4q}) refer to all the group-theory as well as momentum- and frequency-dependent factors which do not depend on $N$ and $\lambda$; their details are unimportant for our arguments.  The only important fact for our analysis is that all the vertices are proportional to $N$ when $\lambda$ is held fixed.  Similarly, in that regime, all the propagators (\ref{eeprpms})-(\ref{eeprpmm}) are proportional to $1/N$.

We summarize the rules for building consistent Feynman diagrams:

\begin{itemize}
\item Quantum ends of propagators are attached to quantum ends of vertices; 
\item Classical ends of propagators are attached to classical ends of vertices; 
\item The following rule is a simple consequence of causality:  If there is a closed loop consisting of a sequence of only $G_A$ (or only $G_R$) propagators, the diagram is identically zero and will be systematically ignored.%
\footnote{Strictly speaking, such diagrams are not illegal, but since they identically vanish, leaving them systematically out will significantly reduce the number of diagrams that need to be drawn for any process.  Also, we do not expect that such diagrams should be independently reproduced on the string-theory side of the duality between the large-$N$ theory and string theory.}
Note that in order for the diagram to be identically zero, the closed loop in question does \textit{not} have to surround just one plaquette.
\end{itemize}

\subsection{Signpost notation for the Feynman diagrams}

Before we proceed to the analysis of the large-$N$ expansion, we find it convenient to introduce a slightly different graphical notation for the non-equilibrium Feynman diagrams, which will simplify the look of the diagrams and allow us to develop some useful intupdflateition.  This new notation will also simplify our proofs and other arguments below.  

Recall that in the $\pm$ formalism, it was very convenient that the ribbon diagrams looked just like those in equilibrium, with all the additional information carried solely by the vertices:  Each vertex was labeled by a sign choice $\pm$.  The type of propagator connecting two given vertices was then uniquely determined by the signs at the vertices.  At first glance, the Feynman rules after the Keldysh rotation do not share the same simplicity.  We will develop a more useful prescription in several simple steps.  First, it is rather awkward to deal with dotted versus undotted halfs of propagators -- we will encode the same information by using regular undotted lines for all the ribbon edges, but placing a ``bulk arrow'' in the middle of the ribbon propagator pointing in the direction from the quantum end to the classical end of the propagator.  The classical-to-classical propagators $G_K$ do not get any ``bulk arrow'' mark.  In the next step, one can pull each arrow from the middle of the propagator to the quantum end of the propagator, and associate this arrow with the adjacent vertex instead.  In this way, each vertex is uniquely assigned a collection of arrows rooted at the vertex and pointing in the directions of various attached propagators.  This collection of arrows rooted at the same vertex is reminiscent of a signpost at trail intersections.  For the lack of a better term, from now on we will refer to this collection of arrows at a given vertex as a ``signpost'', and this notation as the ``signpost notation''.%
\footnote{A somewhat similar notation, using arrows to indicate the $G_A$ and $G_R$ propagators, has been used in the literature (for example see \cite{vilkovisky}).  The novelty of our signpost notation is that we assign the arrows to the vertices, not the propagators.  Note also the different status of the edge arrows reflecting the $SU(N)$ group structure, and the signpost arrows reflecting the non-equilibrium ingredients in our diagrams:  We often indicate graphically only one edge arrow per each closed edge loop or each open edge segment in a given diagram, and do not put edge arrows on all individual propagators, to avoid notational clutter in the figures. On the other hand, the location and direction of each signpost arrow carries important information and such arrows cannot be conveniently left out.}

\begin{figure}[t!]
  \centering
    \includegraphics[width=0.25\textwidth]{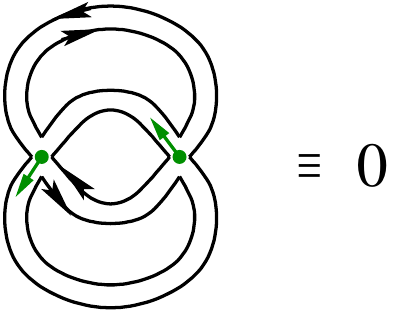}
    \caption{An example of a signpost ribbon diagram which vanishes identically.  Note that in this example, the closed path made of $G_{A}$ propagators that makes this diagram vanish is not surrounding just one plaquette.}
\label{ffrsk13}
\end{figure}

In this new signpost notation, our three-point vertices (\ref{eev3c}, \ref{eev3q})
look as follows, 
\bea
\vcenter{\hbox{\includegraphics[width=.75in]{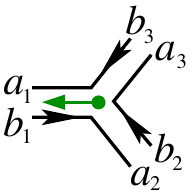}}}\ \ \ &,&\\
\vcenter{\hbox{\includegraphics[width=.75in]{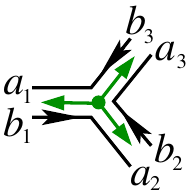}}}\ \ \ &,&\\
\eea
while the four-point vertices (\ref{eev4c}, \ref{eev4q}) are 
\bea
\vcenter{\hbox{\includegraphics[width=.8in]{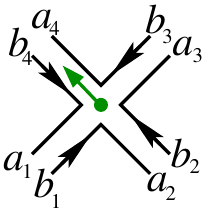}}}\ \ \ &,&\\
\vcenter{\hbox{\includegraphics[width=.8in]{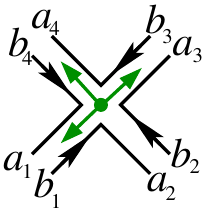}}}\ \ \ &,&\\
\vdots\qquad\quad\ \  &.&\nonumber
\eea
Note that the propagators do not require any additional notation -- each propagator is uniquely determined by the two signposts at the vertices it connects.  The rules for building consistent diagrams can now be rewritten solely as restrictions on the signposts allowed at the vertices of the ribbon diagrams:
\begin{itemize}
\item At each vertex, the signpost carries an odd number of arrows, each pointing into a distinct propagator.
\item The signposts are such that each propagator can have at most one arrow pointing into it from the adjacent vertices.
\item Starting from any vertex, follow the signpost instructions:  Follow any of the adjacent propagators which has an arrow pointing into it; repeat this process at each vertex you visit.  If after $n$ such steps you return to the vertex you started from, the diagram is identically zero and will be systematically omitted.  (See an example in Fig.~\ref{ffrsk13}.) This rule is the rephrasing of the analogous rule we encoutered above in the original notation. 
\end{itemize}

It will be useful to formalize the prescription for traveling along a ribbon diagram $\Delta$ in the direction of the arrows, as follows:  We define an \textit{admissible path} on $\Delta$ from a vertex $v_1$ to another vertex $v_2$ to be a collection of consecutive propagators and vertices, obtained by starting at $v_1$, choosing an arrow from the signpost at $v_1$, moving in the direction of this chosen arrow along the attached propagator to the next vertex, and repeating the steps at each signpost encountered along the way, until we reach $v_2$.  As a consequence of this definition, there is always at least one admissible path going through any given vertex of $\Delta$.  Also, for any pair $v_1,v_2$ of vertices, there might be one or more distinct admissible paths connecting them, or none at all.  

\subsection{All vacuum diagrams vanish identically}

To practice the use of our new notation and to show its efficiency, we will now prove that all vacuum diagrams are zero.  Begin at any vertex, and imagine being a traveler who follows the arrows at all signposts, \ie , travels along an \textit{admissible path} as defined above.  Since the number of arrows at each vertex is odd, there is at least one arrow at your original location.  Follow that arrow, and repeat the step at each new vertex you visit.  Again, since there is at least one arrow at each vertex, this procedure makes sense at each step.  If after a finite number of steps you return to a vertex you already visited, by our rules the diagram is declared to be zero identically.  Since for a vacuum diagram, there are no external legs at which you could end up after a finite number of steps, to prevent the diagram from being zero you would have to travel forever, visiting an infinite number of new vertices.  Since in our analysis we only consider Feynman diagrams with a finite number of vertices, this concludes the proof.

Thus, we reach our first conclusion about the universal structure of non-equilibrium string perturbation theory in the Keldysh-rotated form:
\be
\CZ=\sum_{h=0}^\infty \left(\frac{1}{N}\right)^{2h-2}\CF_h(\lambda,\ldots)=0;
\ee
the sum of all the 0-point diagrams vanishes identically.%
\footnote{Note that as in \cite{neq} and \cite{ssk}, we continue denoting the sum over all \textit{connected} ribbon diagrams by $\CZ$.  As usual, the sum $\SZ$ over \textit{all} diagrams, connected or not, is related to $\CZ$ by $\CZ=\log\SZ$.}
This is an example of the efficiency of the Keldysh-rotated formalism, which must be reproduced by any candidate for the description of the worldsheet dynamics of the string.

\section{Large \textit{N} and string worldsheets:  Classical and quantum surfaces}

We are now ready to demonstrate that for each ribbon diagram $\Delta$ in the Keldysh-rotated formalism, its associated Riemann surface $\Sigma(\Delta)$ can be naturally decomposed into a classical part $\Sigma^\cl$ plus its quantum ``embellishment'' part $\Sigma^\qu$.  This will be done in two steps: First, we define for each diagram its ``classical foundation'' $\hat\Sigma^\cl$:  a surface whose topology is generally simpler (or at least not more complicated) than that of $\Sigma$.  The full surface $\Sigma$ is then obtained by replacing a collection of non-overlapping disks on $\hat\Sigma^\qu$ with the quantum ``embellishments''.  However, since we have just shown that all vacuum diagrams vanish, we cannot use vacuum diagrams to illustrate our arguments as we did in the $\pm$ representation \cite{neq} -- we will need $n$-point correlation functions.

\subsection{Adding external sources}

In what follows, we mostly concentrate for simplicity on diagrams which contribute to the two-point correlator of $\SM$ and $M$,
\be
\left\langle\SM^a{}_bM^c{}_d\right\rangle=\left\langle\SM^a{}_bM^c{}_d\right\rangle_0+\ldots,
\ee
an equation represented graphically as follows, 
\be
\vcenter{\hbox{\includegraphics[width=1.1in]{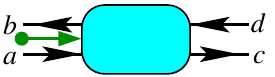}}}\ \ =\ \ \vcenter{\hbox{\includegraphics[width=1.1in]{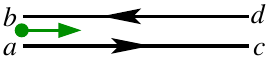}}}\ \ +\ \ldots ,
\ee
where the ``$\ldots$'' denote all the loop corrections.  In fact, in order to eliminate the loose indices at the ends of the propagators, it will be better to couple $M^a{}_b$ and $\SM^a{}_b$ to their conjugate sources, $J^b{}_a$ and $\SJ^b{}_a$, and encode the two-point function (and all higher $n$-point functions) in $SU(N)$ singlets such as
\be
J^b{}_a\left\langle\SM^a{}_bM^c{}_d\right\rangle\SJ^d{}_c.
\label{eejmmj}
\ee
Note that it is the \textit{classical} source $J$ that couples to the \textit{quantum} field $\SM$, and the \textit{quantum} source $\SJ$ to the \textit{classical} field $M$.  This follows from the fact that in the $\pm$ formalism, the coupling to sources adds the following term to the full action (\ref{eeska}), 
\be
\int dt\,\Tr\left(J_+M_+-J_-M_-\right).
\label{eejmpm}
\ee
With the standard definitions
\bea
J_\cl&\equiv&J=\frac{1}{2}(J_++J_-),\nonumber\\
J_\qu&\equiv&\SJ =J_+-J_-,
\eea
the coupling in (\ref{eejmpm}) indeed adds to the Keldysh-rotated action (\ref{eeactk}) the following source term,
$$
\int dt\,\Tr\,(J\SM+\SJ M).
$$

In our ribbon diagrams, we will graphically denote the external sources as follows,
\bea
J:\ \ \ &&\ 
\vcenter{\hbox{\includegraphics[width=.65in]{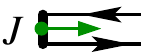}}}\ \ , \nonumber\\
\SJ:\ \ \ &&\ 
\vcenter{\hbox{\includegraphics[width=.65in]{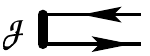}}}\ \ .\nonumber
\eea
Using this notation, the expression in (\ref{eejmmj}) is graphically represented by
\be
\vcenter{\hbox{\includegraphics[width=1.3in]{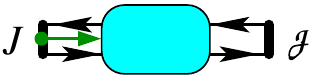}}}.
\ee
On the string dual side, the insertions of the singlets $J\SM$ and $\SJ M$ will correspond to marked points on the surface $\Sigma$.  

In the full theory, one is more appropriately interested in correlation functions of $n$ local composite operators $\CO_i(M,\SM)$ (with $i=1,\ldots, n$) that are \textit{singlets} of the $SU(N)$ symmetry.  It is such operators that can be expected to be associated with simple local vertex-operator insertions on the worldsheets in the dual string theory. The sources $J$ and $\SJ$ that we use to form the singlets $J\SM$ and $\SJ M$ can be simply seen as placeholders for the insertions of such more complicated singlet operators $\CO_i$, and we use them solely for the convenience of our presentation. 

\subsection{Reduction of $\Sigma$ to its classical foundation $\hat\Sigma^\cl$}
\label{ssclfound}

Each ribbon diagram $\Delta$ can be associated with a unique surface $\Sigma$, constructed by simply forgetting the non-equilibrium signposts at the vertices and following the prescription for $\Sigma$ that worked in equilibrium.  Restoring the signposts will then equip $\Sigma$ with some additional structure, and we expect the topological sum over surfaces of genus $h$ to be correspondingly refined.  

How do we identify the refined structure that is naturally induced on $\Sigma$ by the restoration of the non-equilibrium data?  There is one physically well-motivated decomposition of each non-equilibrium ribbon diagram $\Delta$, which induces a natural decomposition of $\Sigma$. Recall first that the information about the \textit{state} is carried by the Keldysh propagators $G_K$, but not the $G_A$ and $G_R$ propagators and the vertices.  It is then natural to define an operation which acts on a ribbon diagram $\Delta$ by ``forgetting'' the $G_K$ propagators:  Erasing all the $G_K$ propagators from a ribbon diagram should leave a subdiagram $\hat\Delta$, in which the information about the state of the system has been erased.  Note that since every vertex of $\Delta$ has at least one arrow at its signpost, no vertices are erased in the process of ddforming $\hat\Delta$.  Some of the vertices of $\hat\Delta$ will have fewer attached legs than their counterparts in $\Delta$.  In particular, some vertices in the reduced diagram $\hat\Delta$ might become ``1-vertices'' or ``2-vertices,'' but each vertex still has at least one propagator attached to it.  Even to such generalized diagrams, one can still apply the standard process of constructing an associated compact surface without boundaries (by gluing in a disk to fill each closed loop).  We will denote this surface by $\hat\Sigma^\cl$ and refer to as the ``classical foundation'' of $\Sigma$.  By design, the expectation is that even on the string side, the classical foundation $\hat\Sigma^\cl$ should be encoding the information about the dynamics but not about the state of the original system.

Note that $\hat\Sigma^\cl$ is either topologically simpler than $\Sigma$, or at most topologically equivalent to $\Sigma$.  In technical terms, the increasing topological complexity of surfaces is measured by the decreasing value of their Euler number.  It turns out that the Euler number of the classical foundation $\hat\Sigma^\cl$ is always greater than or equal to the Euler number of $\Sigma$.  We postpone the proof of this statement until Section~\ref{ssplaq}, after we define more precisely the decomposition of $\Sigma$ into its classical and quantum part.

\subsection{Topology of the classical foundation $\hat\Sigma^\cl$}

First, we will show that the classical foundations $\hat\Sigma^\cl$ that emerge from consistent diagrams can be arbitrarily topologically complicated, as two-dimensional orientable surfaces without boundaries.  Since such two-dimensional surfaces are topologically classified by their non-negative integer genus $n$, we need to show that $\hat\Sigma^\cl$ for all possible $n$ arise from consistent ribbon diagrams.  We will prove this statement by constructing a sequence of ribbon diagrams which contain no $G_K$ propagators, implying that their associated surface $\Sigma$ is identical to its classical foundation, $\hat\Sigma^\cl=\Sigma$, and with $\Sigma$ of arbitrarily high genus.  We will illustrate this on the two-point functions with the $J\SJ$ external source insertions.  

\begin{figure}[b!]
  \centering
    \includegraphics[width=0.36\textwidth]{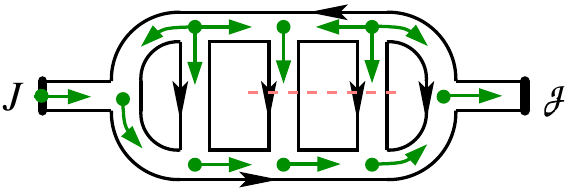}
    \caption{An example of a ribbon diagram without any $G_K$ propagators; thus, the associated surface $\Sigma$ has no quantum embellishments, and $\hat\Sigma^\cl=\Sigma$, the two-pointed sphere.  Cutting the two propagators inside this diagram across the indicated dashed line, and re-gluing them in the opposite order, turns $\Sigma=\Sigma^\cl$ into a two-pointed torus.}
    \label{ffrsk5}
\end{figure}

First, consider the diagram in Fig.~\ref{ffrsk5}.  It is planar, contains only $G_A$ and $G_R$ propagators, and $\Sigma=\hat\Sigma^\cl$ is a two-sphere.  Following the cutting and re-gluing procedure on the two indicated propagators as described in Fig.~\ref{ffrsk5} gives $\Sigma$ which is a two-torus, again isomorphic to $\hat\Sigma^\cl$.  

\begin{figure}[t!]
  \centering
    \includegraphics[width=0.54\textwidth]{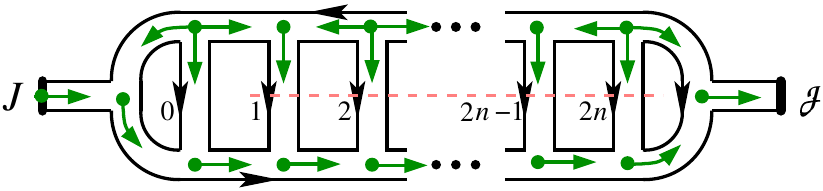}
    \caption{A construction that yields a higher-genus $\Sigma$ with no quantum embellishments.  Starting from this planar diagram, cut the propagators labeled 1 to $2n$ across, along the indicated dashed line, and re-glue them in the opposite order: 1 to $2n$, 2 to $2n-1$, $\ldots$, $2n$ to 1.  This gives a ribbon diagram associated with $\Sigma=\Sigma^\cl$ of genus $n$.}
    \label{ffrsk5a}
\end{figure}

In the next step, we iterate this procedure to form any higher genus surface $\Sigma$, again isomorphic to its classical foundation $\hat\Sigma^\cl$:  Starting with the planar ladder diagram with $2n+1$ rungs as indicated in Fig.~\ref{ffrsk5a}, we cut the $2n$ indicated rungs, and re-glue them in the opposite order.  Counting the number of plaquettes, propagators and vertices of the resulting non-planar diagram demonstrates that its associated surface is of genus $n$.  Since there were no $G_K$ propagators involved, the classical foundation $\hat\Sigma^\cl$ is isomorphic to $\Sigma$, and therefore also of genus $n$.

\subsection{Decomposition of $\Sigma$ into its classical and quantum parts $\Sigma^\cl$ and $\Sigma^\qu$}

When we restore the $G_K$ propagators in a given ribbon diagram $\Delta$, we reconstruct the full surface $\Sigma$ from the classical foundation $\hat\Sigma^\cl$.  This process defines a decomposition of the original surface $\Sigma$ into its quantum and classical parts, which we denote by $\Sigma^\cl$ and $\Sigma^\qu$.  Both $\Sigma^\cl$ and $\Sigma^\qu$ will be two-dimensional surfaces whose boundaries consist of a collection of $S^1$, along which $\Sigma^\cl$ and $\Sigma^\qu$ are glued together.  The classical foundation $\hat\Sigma^\cl$ is then related to $\Sigma^\cl$ simply by gluing in disks into each boundary component of $\Sigma^\cl$.  For an algorithmic definition of this decomposition of $\Sigma$ for any given ribbon diagram $\Delta$, we now refer to a more precise combinatorial description.

\subsection{Combinatorial picture of $\hat\Sigma^\cl$, $\Sigma^\cl$ and $\Sigma^\qu$}
\label{sscomb}

Begin with a ribbon diagram $\Delta$ in the Keldysh-rotated formalism.  The collection of vertices, propagators and closed loops (which we refer to as ``plaquettes'') in $\Delta$ provides a simplicial decomposition of the associated surface $\Sigma$.  We subdivide this combinatorial data associated with $\Delta$ as follows:

\begin{itemize}
  \item
    All vertices belong to $\Sigma^\cl$.
  \item
    All $G_A$ and $G_R$ propagators belong to $\Sigma^\cl$.
  \item
    All plaquettes that have no adjacent $G_K$ propagators belong to $\Sigma^\cl$.
  \item
    All $G_K$ propagators belong to $\Sigma^\qu$.
  \item
    All plaquettes with at least one adjacent $G_K$ propagator belong to $\Sigma^\qu$.
\end{itemize}

This assigns each building block of the cellular decomposition of $\Sigma$ to either $\Sigma^\cl$ or $\Sigma^\qu$.  (Perhaps the only exception is the treatment of the external source insertions, to which we return in Section~\ref{sssour}.)   What is less clear is that $\Sigma^\cl$ and $\Sigma^\qu$ can be naturally interpreted as smooth surfaces, connected to each other along a common boundary which is topologically just a collection of $S^1$'s.  That it is indeed so can be demonstrated by an equivalent definition of the decomposition of $\Sigma$ into $\Sigma^\cl$ and $\Sigma^\qu$, which works plaquette-by-plaquette, and follows a similar plaquette-by-plaquette definition of the triple decomposition of $\Sigma$ in terms of the widened cuts in the $\pm$ formalism studied in detail in \cite{neq,ssk}.

\subsection{Plaquette-by-plaquette construction of a smooth $\Sigma^\qu$}
\label{ssplaq}

\begin{figure}[t!]
  \centering
    \includegraphics[width=0.54\textwidth]{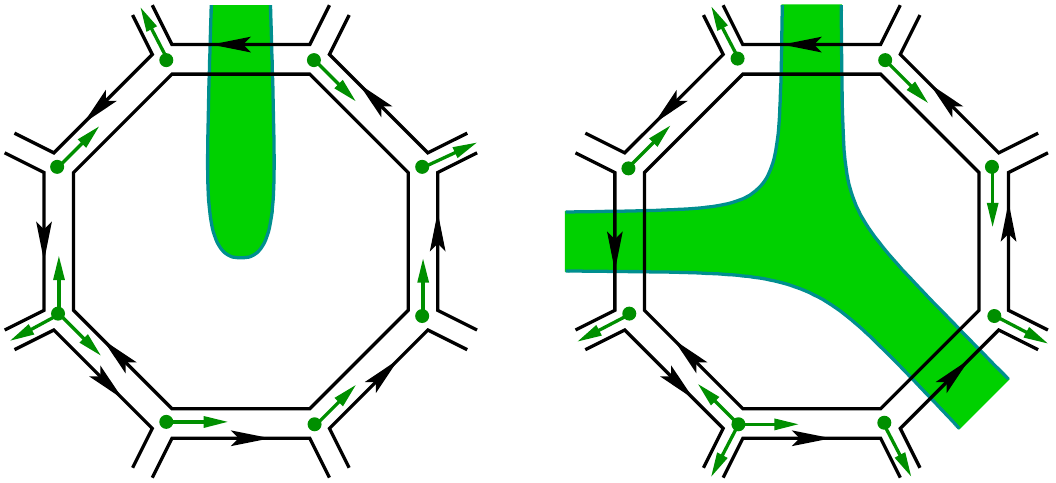}
    \caption{Two examples illustrating the direct plaquette-by-plaquette construction of $\Sigma^\qu$ as a surface with smooth boundaries.   For those readers viewing this figure in color, the portions of $\Sigma^\qu$ so constructed are denoted in green.}
\label{ffrsk12}
\end{figure}

Begin by placing a transverse line segment across the middle of each $G_K$ propagator, and widen this cut into a segment of a two-dimensional ribbon.  Inside each plaquette, connect all such line segments entering the plaquette to the marked center inside the plaquette.  Widening the resulting graph gives a unique portion of a smooth surface with boundaries inside the plauqette, a portion which connects smoothly to other such portions of a smooth surface with boundaries across each adjacent $G_K$ propagator (see Fig.~\ref{ffrsk12} for illustrations).  Their union thus defines a smooth surface with a smooth boundary consisting of a number of $S^1$'s.  It is easy to see that this surface is topologically canonically equivalent to our  $\Sigma^\qu$ as defined combinatorially above.

Another natural perspective on the plaquette-by-plaquette construction is obtained when we switch from the original ribbon diagram $\Delta$ to its dual ribbon diagram $\Delta^\star$.  (We reviewed this duality of ribbon diagrams in Section~2.8 of \cite{neq}, and used it there to study the triple decomposition of $\Sigma$ in the $\pm$ formalism.)  In this dual picture, the widened cut across each $G_K$ propagator in $\Delta$ represents a certain ribbon propagator of $\Delta^\star$, and their connection to the marked center inside a plaquette with at least one adjacent $G_K$ propagator represents a vertex in $\Delta^\star$.  The collection of all such propagators and vertices of $\Delta^\star$ that have been assigned to $\Sigma^\qu$ thus represents a ribbon subdiagram in $\Delta^\star$, and therefore has a natural interpretation as a topologically smooth surface with boundaries.  This surface is precisely the surface $\Sigma^\qu$ that we obtained from the plaquette-by-plaquette prescription.  
    
Given our combinatorial definition of $\Sigma^\cl$ and $\Sigma^\qu$, it is natural to define the following combinatorial Euler numbers associated with the combinatorial ingredients defining the decomposition,
\be
\chi_\cl(\Delta)=V-P_\cl+L_\cl,\qquad\chi_\qu(\Delta)=-P_\qu+L_\qu.
\ee
Here $V$ is the number of vertices in $\Delta$, $P_\qu$ the number of its Keldysh propagators, $P_\cl=P-P_\qu$ the number of its non-Keldysh propagators, $P_\cl$ the number of plaquettes with no adjacent $G_K$ propagators, and $P_\qu$ the number of the plaquettes with at least one $G_K$ propagator.  By repeating the steps used in the $\pm$ formalism in \cite{neq}, it is straightforward to show that these combinatorial Euler numbers reproduce the Euler characteristics of the smooth surfaces $\Sigma^\cl$ and $\Sigma^\qu$,
\be
\chi_\cl(\Delta)=\chi(\Sigma^\cl),\qquad\chi_\qu(\Delta)=\chi(\Sigma^\qu).
\ee
The sum of the two is of course the Euler number of $\Sigma$,  simply given in terms of the number of handles $h$ as $\chi(\Sigma)=2-2h$.

We can now return to the statement we made in Section~\ref{ssclfound}, that the topology of the classical foundation $\hat\Sigma^\cl$ is simpler than that of the full surface $\Sigma$.  The notion of ``topological simplicity'' of a surface $\Sigma$ is quantified by the Euler number $\chi(\Sigma)$:  The simpler the topology of the surface, the greater its Euler number.  We wish to show that
\be
\chi(\hat\Sigma^\cl)\geq\chi(\Sigma).
\label{eegrtr}
\ee
The proof is now simple, because we can rely on the features of the decomposition of each $\Sigma$ as $\Sigma^\cl\cup\Sigma^\qu$.  Consider the connected components of $\Sigma^\qu$, one by one.  Each such component has some number $b$ of boundaries, $b\geq 1$, along which it connects to $\Sigma^\cl$.  Its Euler number is ${}\leq 2-b$.  Replace this connected component with $b$ disks; the Euler number of the replacement is $b$.  Since $b\geq 1$, the Euler number of the replacement is always greater than or equal to the Euler number of the original connected component of $\Sigma^\qu$.  By definition, the classical foundation $\hat\Sigma^\cl$ is obtained from $\Sigma$ by performing this replacement procedure with all connected components of $\Sigma^\qu$.  Using the additivity property of the Euler number, this demonstrates that the Euler number of $\hat\Sigma^\cl$ must be greater than or equal to that of $\Sigma$, thus proving (\ref{eegrtr}).

Note that according to this definition of topological complexity of a surface, we find that the collection of $n$ disconnected spheres is simpler than a collection of $n'$ spheres if $n>n'$.  This is a consequence of our definition of topological complexity that we can live with.

\subsection{Topology of the quantum embellishments $\Sigma^\qu$}
\label{sstopqu}

In our next step, we show that arbitrarily complicated topologies of the quantum embellishment surfaces $\Sigma^\qu$ can appear from consistent ribbon diagrams.  We will prove this statement by constructing a sequence of ribbon diagrams whose quantum embellishments $\Sigma^\qu$ are connected surfaces with one boundary and an arbitrarily high genus.

\begin{figure}[t!]
  \centering
    \includegraphics[width=0.36\textwidth]{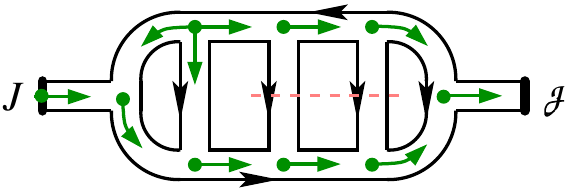}
    \caption{An example of a ribbon diagram with two $G_K$ propagators, whose $\Sigma$ is again a two-pointed sphere, and $\Sigma^\cl=\Sigma$.  In this case, the quantum embellishment $\Sigma^\qu$ is a disk.  Cutting the two indicated propagators along the dashed line and regluing them in the opposite order gives $\Sigma$ which is a two-pointed torus, with $\Sigma^\cl$ a two-pointed sphere, and $\Sigma^\qu$ a torus with one boundary.}
    \label{ffrsk6}
\end{figure}

\begin{figure}[t!]
  \centering
    \includegraphics[width=0.54\textwidth]{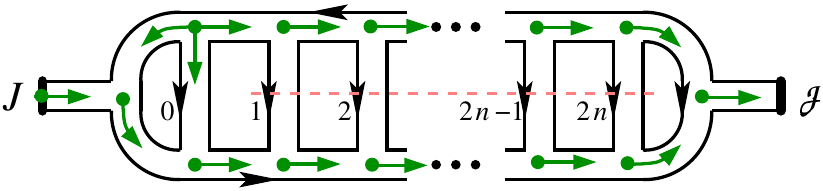}
    \caption{The construction of a surface with a higher-genus quantum embellishment $\Sigma^\qu$.  The indicated diagram gives $\Sigma$ a two-pointed sphere, with $\Sigma^\qu$ a disk, just like in Fig.~\ref{ffrsk6}.  Cutting propagators labeled 1 to $2n$ and regluing them back in the opposite order as in Fig.~\ref{ffrsk5a} yields $\Sigma$ which is a two-pointed surface with $n$ handles, $\hat\Sigma^\cl$ a two-pointed sphere, and $\Sigma^\qu$ with $n$ handles and one boundary.}
    \label{ffrsk6a}
\end{figure}

This constrction is illustrated in Figs.~\ref{ffrsk6} and~\ref{ffrsk6a}:  First, we construct a surface whose classical foundation is an $S^2$ with two marked points at which the sources are inserted, and the quantum part $\Sigma^\qu$ is a torus with one boundary.  Then we iterate this process, and construct a surface whose $\Sigma^\qu$ has any number of handles and one boundary.

\begin{figure}[b!]
  \centering
    \includegraphics[width=0.19\textwidth]{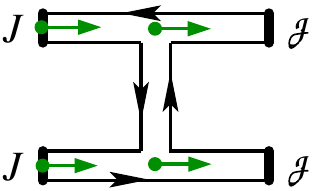}
    \caption{This diagram is planar, and $\Sigma$ is an $S^2$ with four marked points.  The classical foundation $\hat\Sigma^\cl$ consists of two disconnected $S^2$'s, each with two marked points.  The quantum embellishment $\Sigma^\qu$ is a cylinder: Combinatorially, it is constructed from one $G_K$ propagator and the one plaquette of this diagram, and has two boundaries.}
    \label{ffrsk8}
\end{figure}

Next we need to show that $\Sigma^\qu$ can have connected components with more than one boundary component.  Examples of ribbon diagrams with this feature are easy to find if we consider higher $2n$-point correlation functions.  Consider the diagram in Fig.~\ref{ffrsk8}, which contributes to the 4-point function with the external sources $J\SJ J\SJ$.  This diagram is connected and planar, therefore the surface $\Sigma$ associated with it is the sphere (with four marked points corresponding to the insertions of the two $J$'s and two $\SJ$'s.).  Its $\Sigma^\qu$ is a cylinder, and this diagram thus shows that connected components of $\Sigma^\qu$ can have more than one boundary.  

One can also use this 4-point function to find diagrams whose $\Sigma^\qu$ are connected, have two boundaries, and an arbitrary number $h$  of handles:  Simply replace the one $G_K$ propagator in Fig.~\ref{ffrsk8} by $2h+1$ propagators, glue them to the bottom horizontal ribbon in the order from $1$ to $2h+1$, and to the top horizontal ribbon in the reverse order, from $2h+1$ to $1$.  This resulting ribbon diagram will have $2h+1$ propagators, and just one plaquette.  Its $\Sigma^\qu$ is a connected surface with two boundaries and $h$ handles.

\begin{figure}[t!]
  \centering
\includegraphics[width=0.5\textwidth]{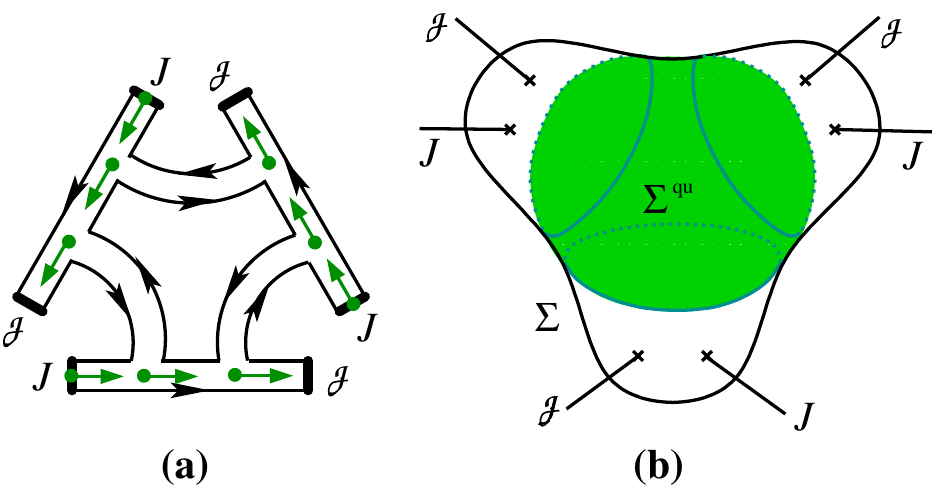}
  \caption{\textbf{(a):} This ribbon diagram is again planar and contributes to a 6-point function.  \textbf{(b):} Its associated surface $\Sigma$ is an $S^2$ with six marked points.  $\hat\Sigma^\cl$ consists of a collection of three $S^2$'s with two marked points each, and $\Sigma^\qu$ is the ``pair of pants'' surface, with no handles and three boundary components.}
    \label{ffrsk9}
\end{figure}

This process can be easily extended to construct examples whose $\Sigma^\qu$ is connected and has more than two boundaries.  One of the simplest ribbon diagrams whose $\Sigma^\qu$ is connected and with three boundaries is depicted in Fig.~\ref{ffrsk9}, and involves a 6-point function.  (In fact, an even simpler ribbon diagram with the same properties would result from removing any one of the three $G_K$ propagators in Fig.~\ref{ffrsk9}.)  Clearly, by iterating this construction to $2b$-point functions, one easily obtains examples whose $\Sigma^\qu$ has $b$ boundary components.

\begin{figure}[b!]
  \centering
    \includegraphics[width=0.53\textwidth]{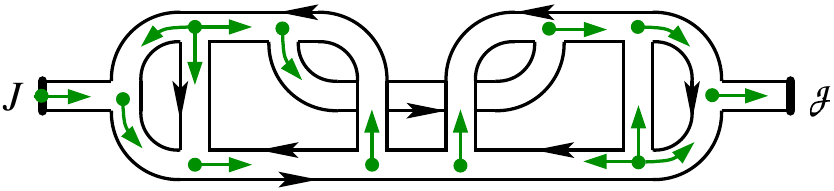}
    \caption{The construction of a surface $\Sigma$ that contributes to the $J\langle\SM\,M\rangle\SJ$ two-point function, and whose $\Sigma^\qu$ is a cylinder.  Here $\Sigma=T^2$, and its classical foundation is $\hat\Sigma^\cl=S^2$.}
    \label{ffrsk7}
\end{figure}

Is it necessary to go to such higher-point functions in order to find examples with connected components of $\Sigma^\qu$ having high numbers $b$ of boundaries, or do such $\Sigma^\qu$ appear already in the 2-point function?  The answer is that they do appear, but in order to find examples of ribbon diagrams that contribute to the 2-point function and whose $\Sigma^\qu$ has a connected component with more than one boundary, one must look a bit harder, to non-planar diagrams.  Consider the diagram in Fig.~\ref{ffrsk7}.  It has been designed such that it only has one $G_K$ propagator.  
It leads to $\Sigma^\qu$ which has two boundary components.  The process clearly iterates, and gives examples of $\Sigma^\qu$ which are connected and have an arbitrary number $b$ of boundary components, even in the case of the 2-point function.  

\subsection{Locations of the external sources}
\label{sssour}

Besides the internal number of vertices, propagators and closed loops, our Feynman diagrams inevitably contain a non-zero number of external source insertions.  The external sources can be either classical $J$, or quantum $\SJ$.  In order to complete the combinatorial rules proposed in Section~\ref{sscomb} to define the decomposition of $\Sigma$ to $\Sigma^\cl$ and $\Sigma^\qu$, we must decide how to assign the external sources to the two parts of this decomposition.

Since the classical source $J$ can never be attached to the $G_K$ propagator, it would appear natural to assign the insertion of $J$ always to $\Sigma^\cl$.  With the quantum source $\SJ$, the story is not so clear:  We can choose to assign it always to $\Sigma^\qu$, or we can choose to assign it to either $\Sigma^\qu$ or $\Sigma^\cl$, depending on whether it is attached to the $G_K$ propagator or the $G_A$ propagator.  Which of these two choices, if any, is more natural?

Perhaps the most natural and elegant answer is to simply admit that the external source insertions are not a part of $\Sigma$, and therefore do not have to be assigned to either $\Sigma^\cl$ or $\Sigma^\qu$.  This picture is further supported by the fact that in critical string theory, the insertions of the vertex operators correspond to the ``punctures'' in the Riemann surface, points which have been removed from $\Sigma$.  This agrees with the observation that each such ``puncture'' contributes $-1$ to the Euler number $\chi(\Sigma)$: In the combinatorial picture, creating a puncture means removing a vertex in the cellular decomposition of $\Sigma$, resulting in the subtraction of 1 from the overall Euler number.  On surfaces with complex structures (such as those in critical string theory in Euclidean worldsheet signature), a puncture can be viewed as an infinitesimally small boundary, and therefore contributes the same amount to $\chi(\Sigma)$.  For $\Sigma$ with $h$ handles, $b$ boundaries and $n$ punctures, the Euler number is then
\be
\chi(\Sigma)=2-2h-b-n.
\ee
This is indeed the expression relevant for the counting of the powers of $N$ in our large-$N$ expansion.  

\begin{figure}[t!]
  \centering
    \includegraphics[width=0.45\textwidth]{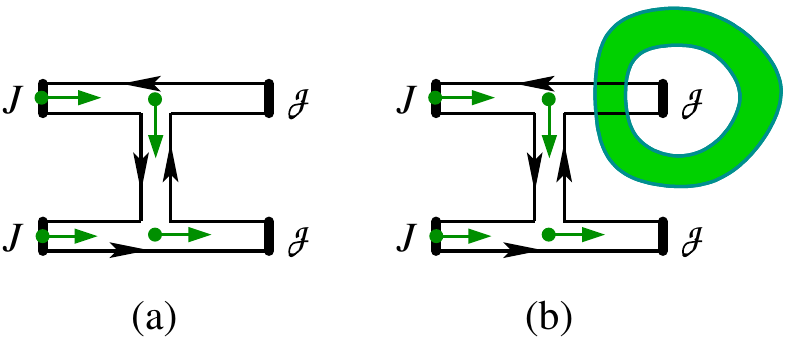}
    \caption{Illustration of a quantum source $\SJ$ attached to a $G_K$ propagator, and its location on $\Sigma$.  \textbf{(a):} This diagram is planar and $\Sigma$ is a four-pointed sphere.  One of the external quantum sources $\SJ$ is connected to a $G_K$ propagator.  \textbf{(b):} Following our rules for the plaquette-by-plaquette construction of $\Sigma^\qu$, we find the decomposition of $\Sigma$ depicted here, with one $\SJ$ isolated inside a disk component of $\Sigma^\cl$.}
    \label{ffrsk11}
\end{figure}

Even if we agree not to consider the source insertions a part of $\Sigma$, a small ambiguity remains:  How do we treat the plaquettes in the ribbon diagram, immediately surrounding the source insertions?  To see a simple example of the possible ambiguity, consider Fig.~\ref{ffrsk11}.  If we follow our plaquette-by-plaquette prescription, one of the quantum sources ends up surrounded by a small disk with no other source insertions, which by our rules is assigned to $\Sigma^\cl$.  This punctured disk is surrounded by $\Sigma^\qu$.  Wouldn't it be more natural and economical to assign this small disk (and its puncture, representing the $\SJ$ insertion) to $\Sigma^\qu$?

\begin{figure}[t!]
  \centering
    \includegraphics[width=0.35\textwidth]{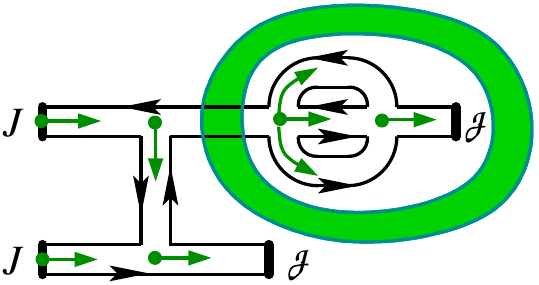}
    \caption{Another ribbon diagram that leads to the same $\Sigma^\cl$ and $\Sigma^\qu$ as the example in Fig.~\ref{ffrsk11}.}
    \label{ffrsk11a}
\end{figure}

We believe that the answer is no, and that the straightforward plaquette-by-plaquette definition of the decomposition is both natural and most economical.  If the diagram in Fig.~\ref{ffrsk11} were the only one with the $\SJ$ insertion surrounded by a disk assigned to $\Sigma^\cl$, it would make sense to re-assign it to $\Sigma^\qu$ and end up with a simplified sum over surface decompositions.  However, there is an entire family of diagrams with the same decomposition into $\Sigma^\cl$ and $\Sigma^\qu$, of which the example in Fig.~\ref{ffrsk11} is only the lowest-order representative, with fewest vertices inside this punctured disk.  Another example is given in Fig.~\ref{ffrsk11a}.  The existence of such higher-order diagrams suggests that it is natural to follow the simple rules of our plaquette-by-plaquette definition of the decomposition of $\Sigma$:  According to that definition, every internal vertex in a given ribbon diagam is always inside $\Sigma^\cl$ -- there is an open disk in $\Sigma$ which contains the vertex and is entirely in $\Sigma^\cl$.  It is then natural to extend this picture also to the 1-vertices associated with the vertex insertions:  Even if the puncture in $\Sigma$ that corresponds to the source insertion is technically not a part of $\Sigma$, it has a neigborhood in $\Sigma$ with the topology of a punctured disk, which intersects only one propagator of the ribbon diagram.  The logic of the plaquette-by-plaquette construction suggests that this punctured disk should be assigned to $\Sigma^\cl$.  

Thus, the extension of the plaquette-by-plaquette construction to the ribbon diagrams with external source insertions suggests that all insertions of both $J$ and $\SJ$ should be naturally interpreted as punctures in $\Sigma^\cl$.  This is the definition of the decomposition of $\Sigma$ with punctures into its classical and quantum part which we adopt for the rest of this paper:  All punctures of $\Sigma$ will always belong to $\Sigma^\cl$.

\section{Non-equilibrium string perturbation theory after the Keldysh rotation}

Thus, we arrive at the form of the topological genus expansion in non-equilibrium string perturbation theory, in the Keldysh-rotated form.  Consider again the sum over all \textit{connected} ribbon diagrams in our generic large-$N$ non-equilibrium system with matrix degrees of freedom, and with $n_\cl$ insertions of the classical source $J$ and $n_\qu$ insertions of the quantum source $\SJ$.  This amplitude can be written as
\be
\CA_{n_\cl,n_\qu}(N,\lambda,\ldots)\,J^{n_\cl}\SJ^{n_\qu}.
\label{eedefa}
\ee
The coefficients $\CA_{n_\cl,n_\qu}(N,\lambda,\ldots)$ can then be expanded in the powers of $1/N$, leading to the string dual description as a sum over connected worldsheet topologies, each with $n_\cl+n_\qu$ punctures.  

For notational simplicity, we introduce the generating functional $\CZ(J,\SJ)$ of the amplitudes (\ref{eedefa}), defined as a formal sum of (\ref{eedefa}) over all $n_\cl$ and $n_\qu$, and refer to $\CZ(J,\SJ)$ as the ``partition function'' for short.  In this language, we can now summarize the main results of this paper as follows:  The large-$N$ expansion of the partition function for the non-equilibrium system in the Keldysh-rotated version of the Schwinger-Keldysh formalism takes the form of a sum over surface topologies, refined to
\be
\CZ(J,\SJ)=\sum_{h=0}^\infty \left(\frac{1}{N}\right)^{2h-2}\sum_{\substack{\textrm{double decompositions}\\ \chi(\Sigma^\cl)+\chi(\Sigma^\qu)=2-2h}} \CF_{\Sigma^\cl,\Sigma^\qu}(J,\SJ;\lambda,\ldots).
\label{eemain}
\ee
In this non-equilibrium case, the sum over the surface topologies goes over all double decompositions of $\Sigma$ into $\Sigma^\cl$ and $\Sigma^\qu$, such that $\Sigma$ is the connected surface of genus $h$, and with $n_\cl+n_\qu$ marked points inside $\Sigma^\cl$ corresponding to the insertions of $n_\cl$  classical sources $J$ and $n_\qu$ quantum sources $\SJ$.

\begin{figure}[t!]
  \centering
    \includegraphics[width=0.38\textwidth]{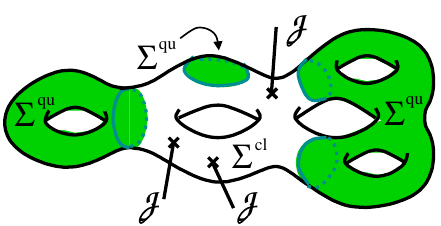}
    \caption{A typical surface $\Sigma$ contributing to (\ref{eemain}), and its decomposition into the classical foundation $\hat\Sigma^\cl$ and the quantum embellishment $\Sigma^\qu$.  In this example, $\Sigma$ is a surface with five handles, and with three $\SJ$ sources inserted at three marked points.  Its classical foundation $\hat\Sigma^\cl$ is a torus with three marked points, and its $\Sigma^\qu$ consists of three disconnected components: A torus with one boundary, a surface with two handles and two boundaries, and a disk.}
    \label{ffrsk10}
\end{figure}

We have already demonstrated that $\CZ(0,0)=0$ identically.  In fact, this observation can be extended from the vacuum diagrams to the more general case of all diagrams with non-zero $J$ but zero $\SJ$,
\be
\CZ(J,0)=0.
\label{eevanj}
\ee
The proof is simple:  Consider a ribbon diagram with at least one vertex.  There is at least one arrow at the signpost at that vertex.  Follow any admissible path starting in the direction of this arrow.  In a diagram with a finite number of vertices, this path must end in a finite number of steps.  The only place where an admissible path can end is at a $\SJ$ source insertion.  Thus, for the diagram to be non-zero, there must be at least one $\SJ$ attached.  There are no diagrams that would contribute to a correlation function with $n$ classical sources $J$, if there is not at least one $\SJ$ source insertion, thus proving (\ref{eevanj}).  Of course, this proof is perturbative in nature, as are all our arguments based on the perturbative expansion in terms of the underlying perturbative ribbon diagrams.  

These vanishing identities have a clear physical interpretation familiar from the field-theory side of the non-equilibrium system:  Setting the quantum source $\SJ$ to zero is equivalent in the original $\pm$ formalism to setting the sources $J_+$ and $J_-$ on the $C_\pm$ parts of the Schwinger-Keldysh contour equal to each other.  When this is done, the probe of the system by $J_+$ on the forward branch is exactly undone by the compensating probe by $J_-$ on the return path, and all the diagrams contributing to such a process are identically zero.  On the string side, this is reflected by the statement of (\ref{eevanj}):  All contributions from the worldsheets $\Sigma$ without at least one $\SJ$ insertion vanish identically.

In fact, this statement about non-equilibrium string perturbation expansion can be further refined:  For the amplitude associated with a given decomposition of $\Sigma$ into $\Sigma^\cl$ and $\Sigma^\qu$ to be non-zero, each connected component of $\Sigma^\cl$ must have at least one $\SJ$ insertion.  The proof is a simple generalization of the argument we used to prove (\ref{eevanj}):  Each connected component of $\Sigma^\cl$ has at least one vertex.  There is at least one allowed path that begins at this vertex.  This allowed path stays within the same connected component of $\Sigma^\cl$, and it has to end somewhere after a finite number of steps.  Since it can only end at an $\SJ$ insertion, each connected component of $\Sigma^\cl$ must have at least one such insertion. 

Note that (\ref{eevanj}) can be interpreted as the boundary condition for solving the full generating functional of the correlation functions (\ref{eemain}).  Finding that (\ref{eevanj}) is valid represents an important check of self-consistency for any $\CZ(J,\SJ)$ in non-equilibrium string theory.

\subsection{Resummation of the perturbative expansion}

The decomposition of the string worldsheet $\Sigma$ into its classical and quantum parts suggests a reorganization of the perturbative expansion in string theory:  We can first perform the sum over the topologically inequivalent classical foundations $\hat\Sigma^\cl$, and then sum over all quantum embellishments that can be added to a given $\hat\Sigma^\cl$.  This resummation leads to the following expression, equivalent to (\ref{eemain}):
\be
\CZ(J,\SJ)=\sum_{\hat\Sigma^\cl}\left(\frac{1}{N}\right)^{-\chi(\hat\Sigma^\cl)}\left\{\sum_{b=0}^\infty
\left(\frac{1}{N}\right)^b\left[\sum_{\Sigma^\qu_b}\left(\frac{1}{N}\right)^{-\chi(\Sigma^\qu_b)}\CF_{\hat\Sigma^\cl, b,\Sigma^\qu_b}(J,\SJ;\lambda,\ldots)\right]\right\}.
\label{eeresum}
\ee
Here $\Sigma^\qu_b$ denotes a quantum embellishment surface, not necessarily connected, with $b$ boundary components.
The first sum in (\ref{eeresum}) is over the classical foundations, which are closed surfaces, also not necessarily connected.  
The second sum in (\ref{eeresum}) is over the number $b$ of disks excised in the classical foundation $\hat\Sigma^\cl$, in order to form $\Sigma^\cl$ (and over the distributions of such excisions among the connected components of $\hat\Sigma^\qu$). 
The third sum in (\ref{eeresum}) is over all possible topologically inequivalent quantum embellishment surfaces $\Sigma^\qu$ which have $b$ boundary components, and can therefore be glued to $\Sigma^\cl$ to form the full surface $\Sigma$.  These ingredients are subjected to just one overall constraint: The resulting $\Sigma$ must be connected.

The resummation of the non-equilibrium string perturbation expansion in the form (\ref{eeresum}) exhibits one somewhat unpleasant feature:  For a given classical foundation $\hat\Sigma^\cl$, the sum over inequivalent $\Sigma^\qu$ topologies is not finite, even at a fixed order in the string coupling $1/N$.  This infinity of inequivalent topologies contributing at the same order in $1/N$ for a given $\hat\Sigma^\cl$ has a simple origin:  Disconnected components of $\Sigma^\qu$ with the topology of a disk.  One can excise any number $m$ of disks from $\hat\Sigma^\qu$ and replace them with such disconneted disk components of $\Sigma^\qu$, without changing the Euler number of $\Sigma$ and thus the order in $1/N$ at which this surface contributes to the partition function.

This feature suggests performing yet another resummation: For a given $\hat\Sigma^\cl$, we can split the sum over all quantum embellishments in (\ref{eeresum}) into two steps:  First the sum over any number of connected components of $\Sigma^\qu$ with the disk topology, followed by the sum over all components of $\Sigma^\qu$ whose Euler number is $\leq 0$ (and which are therefore not disks).  For a given classical foundation $\hat\Sigma^\cl$, the first step defines a ``renormalized'' surface obtained by summing over all possible quantum embellishments by disks, and the second sum over topologically nontrivial quantum embellishments at each order in $1/N$ is then a finite sum over finitely many topologically distinct quantum embellishments of the renormalized $\hat\Sigma^\cl$.  

For specific models, or in specific circumstances, it might happen that the sum over quantum embellishments of each connected component of $\hat\Sigma^\cl$ by disks becomes finite.  Indeed, we shall see two such examples in Section~\ref{sscl}, where we consider classical and stochastic limits of the general non-equilibrium quantum systems:  In Section~\ref{ssclass}, we will find an example where all quantum embellishments vanish identically; and in Section~\ref{ssstoch}, we will encounter another example, in which each connected component of $\Sigma^\cl$ can have at most one boundary component, which implies that the sum over its disk embellishments terminates at order one in the number of disks.  

\subsection{Worldsheet decompositions before and after the Keldysh rotation}

We can now compare and contrast the worldsheet decompositions of $\Sigma$ in non-equilibrium string perturbation theory in the original forward-backward formulation and in the formulation after the Keldysh rotation.  

In the $\pm$ formalism, there is a symmetry between the forward and backward parts of the Schwinger-Keldysh contour, which implies a symmetry between the forward and backward parts $\sigp$ and $\sigm$ of the triple decomposition of the worldsheet.  In particular, their combinatorial definitions in terms of the ingredients in the underlying ribbon diagram reflect this symmetry.  The remaining part, $\sigw$, has a different standing:  It represents the part of the worldsheet associated with the instant of time where the forward and backward branches of the Schwinger-Keldysh contour meet.  $\Sigma^\wedge$ does carry its own topological genus expansion, and in this sense it is topologically two-dimensional.  Still, as we discussed in \cite{ssk}, its combinatorial definition suggests that $\sigw$ may be interpreted as geometrically one-dimensional.  

In the Keldysh-rotated formulation, there is no symmetry between the classical and quantum component $\Sigma^\cl$ and $\Sigma^\qu$ of the two-fold decomposition of the worldsheet surface $\Sigma$:  As we saw, the primary ingredient in this decomposition is the classical foundation $\hat\Sigma^\cl$, which is topologically simpler or at most equivalent to $\Sigma$.  Starting with this classical foundation, $\Sigma$ is formed by adding the quantum embellishments represented by $\Sigma^\qu$.  Both $\Sigma^\cl$ and $\Sigma^\qu$ can have topologies of any genus, but there is no similarity or symmetry between them.  

In fact, there appears to be a certain parallel between $\Sigma^\cl$ of the Keldysh-rotated formalism, and the $\sigp$ and $\sigm$ components of the $\pm$ formalism of \cite{neq}.  Analogously, the quantum part $\Sigma^\qu$ in the Keldysh-rotated formalism is somewhat reminiscent of the wedge region $\sigw$ of the $\pm$ formalism.  Indeed, note an intriguing similarity between the combinatorial definition of $\Sigma^\qu$ in the Keldysh-rotated formalism as given in Section~\ref{sscomb}, and the worldsheet region $\Sigma^\wedge$ at the ``end of time'' in the $\pm$ formalism of \cite{neq}:  In both instances, these surfaces are built solely from propagators and plaquettes, and no vertices in the original ribbon diagram.  Thus, in the Poincar\'e dual ribbon diagram, $\Sigma^\qu$ and $\Sigma^\wedge$ are both built from vertices and lines only, which can make them appear geometrically one-dimensional.  Yet, topologically they correspond to two-dimensional surfaces and carry their own genus expansion, as we demonstrated in Section~\ref{sstopqu}.

\section{Classical limits of non-equilibrium systems and string theory}
\label{sscl}

In non-equilibrium theory in the Keldysh form, there are several popular approximations, which represent various classical limits of the system.  In this section, we study the consequences of taking such limits for the string perturbation expansion.  Besides the interst in studying the string-theory side of such approximations for their own sake, this section serves one additional purpose:  We will see that our results will give further justification to our terminology, and in particular clarify why it makes sense to refer the two parts $\Sigma^\cl$ and $\Sigma^\qu$ as the ``classical'' and ``quantum'' part of the worldsheet $\Sigma$.

\subsection{The classical limit}
\label{ssclass}

The first popular approximation is one in which we consider the quantum field $\phi_\qu$ (or, in our matrix case, $\SM$) to be small compared to $\phi_\cl$ (in our case $M$), expand the action up to linear order in the quantum field $\SM$ and then integrate $\SM$ out (see, \eg , \cite{kamenev,kamenevlesh,kamenevre,mson}).
\be
S_{\textrm{SK}}=\frac{1}{g^2}\int dt\,\Tr\left(M\,G_R^{-1}\SM+\SM\,G_A^{-1}\,M+3\,M^2\SM+4\,M^3\SM+\ldots \right).
\label{eeactkcc}
\ee
Integrating $\SM$ yields a delta function, which makes the remaining dynamical field $M$ satisfy its classical equation of motion,
\be
\ddot M(t)=-V'(M(t)).
\ee
(Here $V$ is the potential that contains all the cubic and higher interaction terms of the original action, and we have kept all the spatial-momentum dependence in the equation implicit.)  Thus, in this limit, all fluctuations (both quantum and thermal) are infinitely suppressed.  This is the reason why this approximation is usually invoked to justify the terminology ``classical'' and ``quantum'' for the fields $M$ and $\SM$:  The ``classical'' field $M$ in this ``classical'' approximation satisfies the classical equation of motion, and the ``quantum'' field has been integrated out.  

What does this approximation look like in the string-theory representation?
Consider the general ribbon diagrams in this approximation.  First, linearizing the cubic and higher interaction terms in the action (\ref{eeactk}) in the quantum field $\SM$ means that we drop all vertices with more than one arrow at their signpost.  Note a curious consequence:  In this classical approximation, there is no free will left for our hypothetical travelers following admissible paths on a given ribbon diagram!  Indeed, the choice of an admissible direction at each vertex along the path is uniquely determined by the single arrow at its signpost, and all admissible paths are completely deterministic.

Linearizing the quadratic term in (\ref{eeactk}) means that we keep only the mixed propagators $G_A$ and $G_R$, dropping all the $G_K$ propagators.  This step is familiar:  This is how we defined the reduction from the full surface $\Sigma$ to its classical foundation $\hat\Sigma^\cl$ in Section~\ref{ssclfound}.  However, not all classical foundations of the original theory will appear:  Only those diagrams whose every vertex has just one arrow at its signpost will survive the linearization procedure.

It is now easy to show that all such ribbon diagrams will be collections of trees.  Each of the trees is rooted by one $\SJ$ insertion.  The collection of all allowed paths that end at this $\SJ$ form the branches of the tree.  The deterministic feature of the allowed paths discussed above ensures that there are indeed no closed loops in this tree.  Tree diagrams are planar, and therefore $\Sigma$ is just a collection of spheres.

Thus, we reach a very pleasing conclusion:  In the classical limit of the original non-equilibrium quantum system, the partition function $\CZ(J,\SJ)$ as given by the sum over worldsheet topologies automatically reduces itself to a sum over string worldsheet surfaces with only spherical topologies!   For each term, the number of $S^2$'s is equal to the number of $\SJ$ external insertions.  Moreover, all these surfaces have no quantum embellishments $\Sigma^\qu$, and therefore are equivalent to their classical foundation $\hat\Sigma^\qu$.   In equilibrium closed string theory, summing over only spherical topologies is the hallmark of taking the classical limit.  It is nice to see that taking the classical limit of the non-equilibrium system matches the process of taking the classical limit on the string side as well.  We believe that this result provides some intuitive justification for the terminology we introduced for the decomposition of $\Sigma$ into its ``classical'' and ``quantum'' part $\Sigma^\cl$ and $\Sigma^\qu$.

\subsection{Classical stochastic limit and the Martin-Siggia-Rose method}
\label{ssstoch}

In this approximation, we take the semiclassical limit $\hbar\to 0$ but keep the classical thermal fluctuations.  This is achieved by restoring the dependences on $\hbar$ in the Schwinger-Keldysh action (\ref{eeactk}), exposing the system to an environment by coupling it to a thermal bath of harmonic oscillators, and taking the classical limit while keeping the temperature $T$ fixed (see, \eg, \cite{kamenev}, Chapters 3.2 and 4, for details).  Note that this approximation will not require the $M$ degrees of freedom to be in equilibrium, only the bath.

It turns out that keeping the dependence on non-zero $T$ is equivalent to keeping not only the linear terms but also the terms \textit{quadratic} in $\SM$ in our expansion of the action(\ref{eeactk}). The classical action (\ref{eeactkcc}) is then modified to
\be
S_{\textrm{SK}}=\frac{1}{g^2}\int dt\,\Tr\left(M\,G_R^{-1}\SM+\SM\,G_A^{-1}\,M+3\,M^2\SM+4\,M^3\SM+\ldots +i\gamma T\SM^2\right).
\label{eeactkc}
\ee
Here $\gamma$ is a constant that characterizes the spectral density of the Ohmic bath modelling the environment (see \cite{kamenev}).  This constant $\gamma$ also appears in the additive friction terms in the $G_A$ and $G_R$ propagators, terms which were absent in these propagators in the classical limit of Section~\ref{ssclass}; these additional terms do not influence our treatment of the Feynman rules, ribbon diagrams and our conclusions.

In order to see in what sense this action (\ref{eeactkc}) represents a classical stochastic system, it is convenient to use the Hubbard-Stratonovich transformation in the path integral,
\be
e^{-\gamma T\int dt\,\Tr\left(\SM^2\right)}=\int \SD\xi(t)\,e^{-\int dt\,\Tr\left(\frac{1}{\gamma T}\xi^2-2i\xi(t)\SM(t)\right)},
\label{eestoch}
\ee
so that we can trade the term quadratic in $\SM$ for a linear coupling between $\SM$ and a new, typically Gaussian, field $\xi$.  In our case, both $\SM(t)$ and $\xi(t)$ are $SU(N)$ matrices (with all additional dependences on the spatial coordinates or other quantum numbers again kept implicit, as has been the case throughout our analysis).

Since $\SM$ now appears only linearly, it can be again integrated out to give a delta function localized on the stochastic classical equation of motion for $M$,
\be
\ddot M(t)=-\gamma\dot M -V'(M(t))+\xi(t).
\label{eelange}
\ee
In this classical equation, $\xi(t)$ serves as a stochastic noise, with a Gaussian distribution represented by the path integral (\ref{eestoch}).  Note the presence of the friction term $-\gamma\dot M$, which appears due to the dependence of $G_A$ and $G_R$ on $\gamma$ mentioned above.  The famous Martin-Siggia-Rose method \cite{msr} for dealing with stochastic systems reverses this construction \cite{dedo,janssen}:  It starts with a Langevin equation analogous to (\ref{eelange}), and reintroduces the quantum field $\SM$ to represent the system in the path integral language.  

Now we will use the action (\ref{eeactkc}) of this classical stochastic limit of the original system of matrix degrees of freedom, to see the implications of this approximation on the dual string side.  

First note that in this limit, all vertices in the surviving ribbon diagrams are again allowed to have just one arrow at their signpost, just as in the classical limit discussed in Section~\ref{ssclass}.  In particular, the following conclusions about the classical foundation $\hat\Sigma^\cl$ of the surfaces associated with the surviving ribbon digrams stay the same:
\begin{itemize}
\item
  All admissible paths on the ribbon diagrams are completely deterministic;
\item
  The reduced ribbon diagram that defines the classical foundation $\hat\Sigma^\cl$ is a collection of trees (with one tree per each $\SJ$ source insertion), and its each connected component is therefore planar;
\item
  The classical foundation $\hat\Sigma^\cl$ is either a sphere, or a collection of disconnected spheres (with one $S^2$ for each connected tree component of the associated reduced ribbon diagram);
\item
  The number of connected components $S^2$ of $\hat\Sigma^\cl$ is equal to the number of $\SJ$ source insertions in the diagram.  
\end{itemize}

In contrast to the zero-temperature classical limit studied in Section~\ref{ssclass}, however, there is now a non-zero remnant of the classical-to-classical $G_K$ propagator, due to the presence of the $\SM^2$ term in (\ref{eeactkc}) linear in $T$.  Thus, the worldsheets $\Sigma$ contributing in this stochastic classical limit will still contain quantum embellishments, but their classical foundations will be collections of $S^2$'s.

\begin{figure}[t!]
  \centering
  \includegraphics[width=0.54\textwidth]{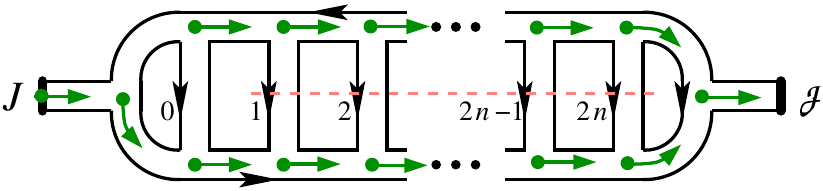}
      \caption{An example a ribbon diagram that follows the rules of the stochastic classical approximation with action (\ref{eeactkc}). Note the deterministic nature of the amissible paths on this diagram.  Cutting across the indicated $2n$ rungs and re-gluing them as in Fig.~\ref{ffrsk6a} yields a ribbon diagram whose surface $\Sigma$ has $n$ handles, while its classical foundation is still $\hat\Sigma^\cl=S^2$.  The non-trivial topology is entirely contained in the quantum part $\Sigma^\qu$, which is a surface with $n$ handles and one boundary.}
    \label{ffrsk6cl}
\end{figure}

Restoring now all the $G_K$ propagators in the reduced diagram, the surface $\Sigma$ can have an arbitrarily high number of handles, as we show in Fig.~\ref{ffrsk6cl}.  However, this nontrivial topology of $\Sigma$ is now solely due to the quantum embellishments:  Leaving out the $G_K$ propagators reduces any original ribbon diagram of this approximation to a collection of trees, implying that the classical foundation $\hat\Sigma^\cl$ is always a collection of two-spheres.  
In addition, one can similarly show that each connected component of $\Sigma^\cl$ is a disk (\ie , it has only one boundary $S^1$ connecting it to $\Sigma^\qu$), or the entire $\Sigma$ is an $S^2$ with no quantum embellishments.  

We see that even in the stochastic classical limit, there continues to be a meaningful sense in which the ``classical'' limit in the non-equilibrium system (as defined by (\ref{eeactkc})) means also a ``classical'' limit in the sense of the dual string theory, where the ``classical'' string limit is conventionally understood as the summation over worldsheets with only spherical topology \cite{gsw} and possibly with marked points.  This time, however, this classical string limit applies only to the \textit{classical foundation} of $\Sigma$, while the higher-genus quantum embellishments $\Sigma^\qu$ represent the classical thermal or stochastic fluctuations in the original matrix system. 

\section{Conclusions}

In this paper we found that, as anticipated, the calculus of non-equilibrium string perturbation theory looks quite different in the Keldysh representation, in comparison to its form in the original $\pm$ formalism that we found in \cite{neq}.  In both cases, the large-$N$ expansion is organized into a sum over surfaces $\Sigma$ of increasing topological complexity, just as in the standard string perturbation theory at equilibrium.  In contrast to equilibrium, however, in both representations of non-equilibrium string perturbation theory the surfaces $\Sigma$ are found to carry a more refined structure (besides just the genus of $\Sigma$) which is universal for all systems.  It is this additional structure that is quite different between the two non-equilibrium representations.

In the $\pm$ formalism, the worldsheet surfaces $\Sigma$ exhibit a triple decomposition, into their forward branch $\sigp$, a backward branch $\sigm$ and the ``wedge'' region $\sigw$ which corresponds to the crossing from the forward to the backward portion of the Schwinger-Keldysh time contour.  In contrast, in the Keldysh representation, each surface $\Sigma$ consists of a classical foundation $\hat\Sigma^\cl$, which is further decorated by the quantum portion $\Sigma^\qu$ of the surface.

In \cite{neq}, we also studied the structure of non-equilibrium string perturbation theory, and the refinement of the worldsheet decompositions, for closed time contours with more than two segments, most notably for the Kadanoff-Baym contour relevant for systems at finite temperature.  Besides the forward branch $C_+$ and the backward branch $C_-$, this time contour has a third segment $\C_\rmm$ (sometimes called the ``Matsubara'' segment), which extends along the imaginary direction by the amount $\beta=1/T$ set by the temperature.  We have not generalized the results of the Keldysh rotation to this case, simply because the status of this third segment is different than that of $C_\pm$.  However, one can certainly imagine a hybrid formalism, in which the Keldysh rotation has been performed on the fields taking values on $C_\pm$, leaving the Matsubara segment intact.  Such a hybrid formalism has indeed been used extensively in the theory of non-equilibrium many-body systems (see \cite{spicka} for a review). The fields in this hybrid formalism would consist of the classical and quantum fields $M(t)$ and $\SM(t)$ that we studied in this paper, plus the Matsubara field $M_\rmm(\tau)$ that we used in Section~3 of \cite{neq}.  By combining the results of \cite{neq} and those of the present paper, it should be possible to derive the form of the worldsheet decomposition in this hybrid formalism for non-equilibrium systems with a string dual.

Effectively, our analysis in \cite{neq} and in the present paper produced a set of rules which can be viewed almost as axioms, and which are so universal that we expect any string theory out of equilibrium to be consistent with them:  In the $\pm$ description, the instant in time where the forward and backward contours meet is perceived from the worldsheet perspective as topologically two-dimensional, and carries its own genus expansion; the sum over surfaces is refined into a sum over their triple decompositions.  In the Keldysh rotated description, each part of the two-fold decomposition of the worldsheet surface into its classical foundation and quantum embellishments carries its own independent genus expansion.  Due to their universal nature, these axioms are arguably not very strong, and therefore not very helpful in determining any specific details of the worldsheet dynamics.  We hope, however, that they may at least provide some guidance in the future search for the worldsheet description, in particular examples of interest.

It will be interesting to see which of the two representations of the string-theoretic dual description of large-$N$ non-equilibrium systems will be more useful from the perspective of the worldsheet theory.  Perhaps the answer might even depend on the large-$N$ system in question, and the kind of string theory which happens to be dual to it.  We leave these fascinating questions open for future investigations.

\acknowledgments
We wish to thank Andr\'es Franco Valiente for useful comments on this manuscript.
This work has been supported by NSF grant PHY-1820912.  

\bibliographystyle{JHEP}
\bibliography{neq}

\providecommand{\href}[2]{#2}\begingroup\raggedright\begin{thebibliography}{10}

\bibitem{neq}
P.~Ho\v{r}ava and C.~J. Mogni, \emph{{Large-$N$ expansion and string theory out
  of equilibrium}},  \href{https://arxiv.org/abs/arXiv:2008.11685}{{\ttfamily
  arXiv:2008.11685}}.

\bibitem{vilkovisky}
G.~Vilkovisky, \emph{{Expectation values and vacuum currents of quantum
  fields}}, {\emph{Lect. Notes Phys.} {\bfseries 737} (2008) 729}
  [\href{https://arxiv.org/abs/arXiv:0712.3379}{{\ttfamily arXiv:0712.3379}}].

\bibitem{rammer}
J.~Rammer, \emph{{Quantum field theory of non-equilibrium states}}. Cambridge
  University Press, Cambridge, 2007.

\bibitem{kamenev}
A.~Kamenev, \emph{{Field Theory of Non-Equilibrium Systems}}. Cambridge
  University Press, Cambridge, 2011.

\bibitem{kamenevlesh}
A.~{Kamenev}, \emph{{Many-body theory of non-equilibrium systems}},
  \href{https://arxiv.org/abs/arXiv:cond-mat/0412296}{{\ttfamily
  arXiv:cond-mat/0412296}}.

\bibitem{kamenevre}
A.~Kamenev and A.~Levchenko, \emph{{Keldysh technique and nonlinear
  sigma-model: Basic principles and applications}},
  \href{https://doi.org/10.1080/00018730902850504}{\emph{Adv. Phys.} {\bfseries
  58} (2009) 197} [\href{https://arxiv.org/abs/arXiv:0901.3586}{{\ttfamily
  arXiv:0901.3586}}].

\bibitem{keldysh}
L.~Keldysh, \emph{{Diagram technique for nonequilibrium processes}}, {\emph{Zh.
  Eksp. Teor. Fiz.} {\bfseries 47} (1964) 1515}.

\bibitem{lo}
A.~Larkin and Y.~Ovchinnikov, \emph{{Nonlinear conductivity of superconductors
  in the mixed state}}, {\emph{JETP} {\bfseries 41} (1975) 960}.

\bibitem{lw}
D.~C. Langreth and J.~W. Wilkins, \emph{{Theory of Spin Resonance in Dilute
  Magnetic Alloys}}, \href{https://doi.org/10.1103/PhysRevB.6.3189}{\emph{Phys.
  Rev. B} {\bfseries 6} (1972) 3189}.

\bibitem{spicka}
V.~\v{S}pi\v{c}ka, B.~Velick\'{y} and A.~Kalvov\'{a}, \emph{{Electron systems
  out of equilibrium: Nonequilibrium Green's function approach}},
  \href{https://doi.org/10.1142/S0217979214300138}{\emph{Int. J. Mod. Phys.}
  {\bfseries B 28} (2014) 1430013}.

\bibitem{langreth}
D.~C. Langreth, \emph{Linear and nonlinear response theory with applications},
  in \emph{Linear and Nonlinear Electron Transport in Solids} (J.~T. Devreese
  and V.~E. van Doren, eds.).
\newblock Plenum Press, 1976.

\bibitem{svl}
G.~Stefanucci and R.~van Leeuwen, \emph{{Nonequilibrium Many-Body Theory of
  Quantum Systems}}. Cambridge University Press, Cambridge, 2013.

\bibitem{ghh}
M.~Gell-Mann and J.~B. Hartle, \emph{{Classical equations for quantum
  systems}}, \href{https://doi.org/10.1103/PhysRevD.47.3345}{\emph{Phys. Rev.
  D} {\bfseries 47} (1993) 3345}
  [\href{https://arxiv.org/abs/arXiv:gr-qc/9210010}{{\ttfamily
  arXiv:gr-qc/9210010}}].

\bibitem{fv}
R.~Feynman and J.~Vernon, F.L., \emph{{The Theory of a general quantum system
  interacting with a linear dissipative system}},
  \href{https://doi.org/10.1016/0003-4916(63)90068-X}{\emph{Annals Phys.}
  {\bfseries 24} (1963) 118}.

\bibitem{fhibbs}
R.~P. Feynman and A.~R. Hibbs, \emph{{Quantum mechanics and path integrals}},
  International series in pure and applied physics. McGraw-Hill, New York, NY,
  1965.

\bibitem{ssk}
P.~Ho\v{r}ava and C.~J. Mogni, \emph{{String perturbation theory on the
  Schwinger-Keldysh time contour}},
  \href{https://arxiv.org/abs/arXiv:2009.03940}{{\ttfamily arXiv:2009.03940}}.

\bibitem{mson}
A.~Mueller and D.~Son, \emph{{On the Equivalence between the Boltzmann equation
  and classical field theory at large occupation numbers}},
  \href{https://doi.org/10.1016/j.physletb.2003.12.047}{\emph{Phys. Lett. B}
  {\bfseries 582} (2004) 279}
  [\href{https://arxiv.org/abs/arXiv:hep-ph/0212198}{{\ttfamily
  arXiv:hep-ph/0212198}}].

\bibitem{msr}
P.~Martin, E.~Siggia and H.~Rose, \emph{{Statistical Dynamics of Classical
  Systems}}, \href{https://doi.org/10.1103/PhysRevA.8.423}{\emph{Phys. Rev. A}
  (1973) 423}.

\bibitem{dedo}
{De Dominicis, C.}, \emph{Techniques de renormalization de la th\'eorie des
  champs et dynamique des ph\'enom\`enes critiques},
  \href{https://doi.org/10.1051/jphyscol:1976138}{\emph{J. Phys. (Paris)
  Colloques} {\bfseries 37} (1976) C1}.

\bibitem{janssen}
H.-K. {Janssen}, \emph{{On a Lagrangean for classical field dynamics and
  renormalization group calculations of dynamical critical properties}},
  \href{https://doi.org/10.1007/BF01316547}{\emph{Zeitschrift fur Physik B
  Condensed Matter} {\bfseries 23} (1976) 377}.

\bibitem{gsw}
M.~B. Green, J.~Schwarz and E.~Witten, \emph{{Superstring Theory. Vol. 1:
  Introduction}}, Cambridge Monographs on Mathematical Physics. Cambridge U.
  Press, 1988.

\end{thebibliography}\endgroup
\end{document}